\documentclass[sigconf]{acmart}
\AtBeginDocument{%
  }

\setcopyright{cc}
\setcctype{by}
\copyrightyear{2026}
\acmYear{2026}
\acmDOI{}
\acmBooktitle{Proceedings of the CHI 2026 Workshop: Ethics at the Front-End: Responsible User-Facing Design for AI Systems (CHI '26), April 13--17, 2026, Barcelona, Spain}
\acmConference[CHI '26]{CHI 2026 Workshop: Ethics at the Front-End}{April 13--17, 2026}{Barcelona, Spain}

\acmISBN{}

\usepackage{booktabs} 
\usepackage{array}
\usepackage{booktabs}

\begin{document}

\title[Front-End Ethics for Sensor-Fused Health Conversational Agents]{Front-End Ethics for Sensor-Fused Health Conversational Agents: An Ethical Design Space for Biometrics}

\author{Hansoo Lee}
\affiliation{%
  \institution{Imperial College London}
  \city{London}
  \country{UK}
}
\affiliation{%
  \institution{Korea Institute of Science and Technology}
  \country{Republic of Korea}
}
\email{h.lee1@imperial.ac.uk}
\email{hansoolee@kist.re.kr}

\author{Rafael A. Calvo}
\affiliation{%
  \institution{Imperial College London}
  \city{London}
  \country{UK}}
\email{r.calvo@imperial.ac.uk}

\renewcommand{\shortauthors}{Lee et al.}

\begin{abstract}
The integration of continuous data from built-in sensors and Large Language Models (LLMs) has fueled a surge of “Sensor-Fused LLM agents” for personal health and well-being support. While recent breakthroughs have demonstrated the technical feasibility of this fusion (e.g., Time-LLM, SensorLLM), research primarily focuses on “Ethical Back-End Design for Generative AI”, concerns such as sensing accuracy, bias mitigation in training data, and multimodal fusion. This leaves a critical gap at the front end, where invisible biometrics are translated into language directly experienced by users. We argue that the “illusion of objectivity” provided by sensor data amplifies the risks of AI hallucinations, potentially turning errors into harmful medical mandates. This paper shifts the focus to “Ethical Front-End Design for AI”, specifically, the ethics of biometric translation. We propose a design space comprising five dimensions: Biometric Disclosure, Monitoring Temporality, Interpretation Framing, AI Stance, and Contestability. We examine how these dimensions interact with context (user- vs. system-initiated) and identify the risk of biofeedback loops. Finally, we propose “Adaptive Disclosure” as a safety guardrail and offer design guidelines to help developers manage fallibility, ensuring that these cutting-edge health agents support, rather than destabilize, user autonomy. 
\end{abstract}




\maketitle

\section{Introduction}
Early health Conversational agents (CAs) were largely rule-based (scripts, decision trees, templates) or relied on limited NLP, and their primary input was self-report entered by users~\cite{laranjo2018conversational}. In that period, the central front-end design goals were to improve the conversational experience itself: empathic interaction and rapport to support sustained use, safety to avoid medical misinformation, and tone management and privacy protection to maintain trust~\cite{laranjo2018conversational, bickmore2005establishing, may2022security}. 

With the rise of LLMs, health CAs can process much more broader natural-language input and generate more sophisticated outputs such as counselling, summarisation, explanation, and coaching. However, LLM-based health CAs uniquely introduce interaction risks, including hallucinations, overconfident expressions, harmful advice, emotional persuasion (pseudo-empathy), and dependence inducing dynamics~\cite{weidinger2021ethical, lee2023benefits, haug2023artificial, singhal2023large, kerasidou2020artificial}. A key concern is that many LLM-based CAs are not sufficiently coupled with users’ continuous, quantitative personal health data (e.g., sensor-based baselines), so responses may rely on generalised or prototypical health knowledge and become persuasive even when they conflict with an individual’s context. In medical settings, “confidently stating a plausible error” can directly change user behaviour and be particularly harmful, which is why improving back-end performance alone cannot guarantee safety~\cite{lee2023benefits, haug2023artificial, singhal2023large}. These concerns also highlight oversight and regulation needs, and risks such as the empathy trap, where empathic-sounding responses can increase overtrust and emotional dependence~\cite{mesko2023imperative, kerasidou2020artificial}.

To challenges have shifted again in the past one to two years. Wearables and sensing advances now enable routine collection of continuous physiological and behavioural signals (e.g., HR/HRV, sleep stages, activity, respiration, temperature), and research has rapidly expanded approaches that align and fuse these time-series signals with LLMs to support prediction, explanation, and dialogue~\cite{jin2024timellm, pmlr-v248-kim24b, li2025sensorllm, yan2025large, ren2025toward, hoque2025toward, cosentino2024towards, khasentino2025personal, fang2024physiollm, abbasian2025conversational, demirel2025using}. Commercial deployments now foreground data-grounded health conversations (e.g., ChatGPT Health, WHOOP Coach, Oura Advisor)~\cite{openai_chatgpt_health_2026, whoop_coach_openai_2023, oura_advisor_2025}. Sensor-fused CAs increasingly read bodily signals (or connected records), translate them into language, and intervene by conversation.

Sensor-fused health CAs change the ethical terrain because the input is the body; while users view text as subjective, sensor data often feel objective. When LLMs translates signals into definitive meanings (e.g., “you are stressed”), reponses inherits this perceived authority, turning uncertain inferences into diagnosis-like facts, reinforcing an “illusion of objectivity”. Coupled with LLM risks (hallucination, overconfidence, persuasion), alerts may trigger anxiety that worsens physiological states, creating nocebo and biofeedback loops~\cite{mesko2023imperative, lee2023benefits, haug2023artificial, singhal2023large, kerasidou2020artificial}. Moreover, empathy-like phrasing can increase overtrust and emotional dependence in health contexts~\cite{mesko2023imperative, kerasidou2020artificial}. For sensor-fused health CAs, front-end ethics is therefore not merely UI aesthetics but can be reframed as designing the ethical act of how translation is experienced~\cite{schiff2020principles, amershi2019guidelines, world2024ethics}. Ethical failure is not limited to “wrong answers”; it can also arise from how outputs are delivered and how that delivery changes users’ emotions and behaviours.

Accordingly, this paper shifts the ethical focus from “a back end that eliminates errors” to “a front end that translates safely under inevitable uncertainty.” We (1) organise the system and interaction landscape of sensor-fused health CAs, (2) propose a five-dimensional spectrum-based ethical design space for biometric translation, and (3) summarise risk-mitigation principles that break harmful biofeedback loops through concrete interface guardrails.

\section{System and Interaction Landscape for Sensor-Fused Health CAs}

\begin{figure*}[t]
  \centering
  \includegraphics[width=\textwidth]{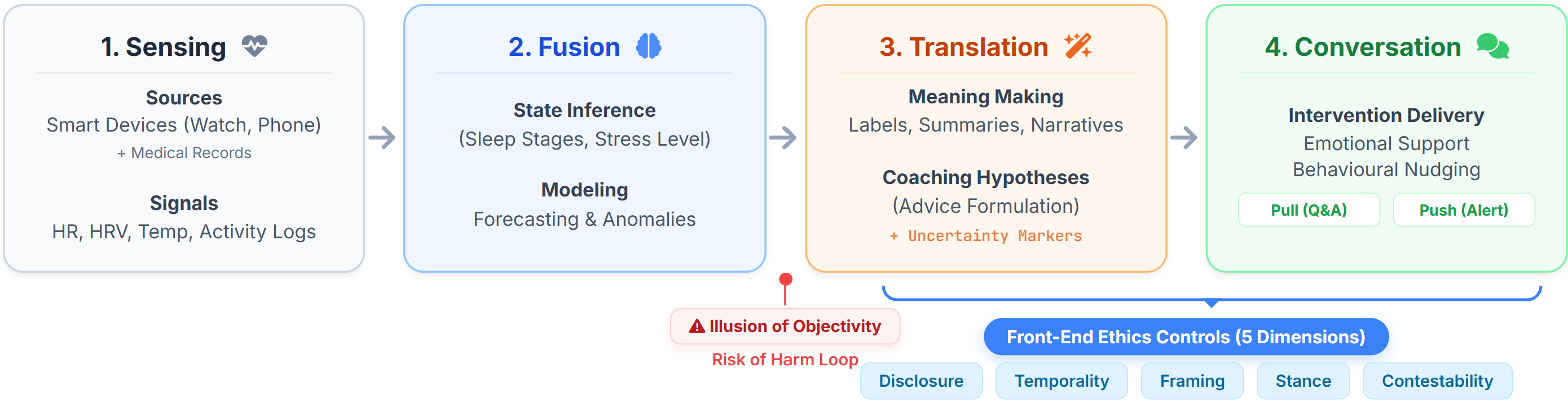}
  \caption{Conceptual architecture and front-end ethics controls in Sensor-Fused Health CAs.}
  \Description{A diagram illustrating the four-stage pipeline (Sensing, Fusion, Translation, Conversation) and showing the 5 Front-End Ethics Controls intervening at the Translation stage to mitigate the 'Risk of Harm Loop'.}
  \label{fig:system_arch}
\end{figure*}

As illustrated in Figure 1, the architecture of sensor-fused health CAs progresses from sensing physiological signals~\cite{song2025predicting} and fusion (state inference)~\cite{jin2024timellm, pmlr-v248-kim24b, li2025sensorllm} to translation and conversation. While engineering efforts often focus on back-end accuracy, front-end ethics is situated at the critical translation boundary, where invisible biometrics are converted into language. Here, user-facing design critically shapes trust calibration and perceived rationale~\cite{amershi2019guidelines, liao2020questioning}. The ethical stakes depend heavily on the interaction trigger. Pull interactions (user-initiated) typically occur when users are ready to interpret data. In contrast, Push interactions (system-initiated alerts) may arrive during moments of busyness, social exposure, or vulnerability. In such contexts, rigid interpretations (e.g., “high stress detected”) can trigger anxiety and worsen physiological states, creating harmful \textit{biofeedback loops}~\cite{kerasidou2020artificial, world2024ethics}. When compounded by LLM risks like hallucinations~\cite{haug2023artificial, lee2023benefits, singhal2023large} and empathy traps~\cite{mesko2023imperative, kerasidou2020artificial}, these loops can severely destabilize user autonomy. Therefore, design controls must focus on managing how translations is experienced.

\section{Ethical Design Space for Biometric Translation}

We propose front-end ethics for biometric translation as five dimensions. These dimensions are design levers whose positions should shift depending on Pull versus Push delivery, service goals, user vulnerability, and risk level. Table~\ref{tab:designspace} summarises the design space and its ethical tensions.

\begin{table*}[tb]
\caption{The Ethical Design Space for Sensor-Fused LLMs}
\label{tab:designspace}
\begin{tabular}{@{}llp{5cm}p{4cm}@{}}
\toprule
\textbf{Dimension}        & \textbf{Definition}                   & \textbf{Spectrum (Left $\leftrightarrow$ Right)}                 & \textbf{Key Ethical Tension} \\ \midrule
1. Data Disclosure        & \textbf{Granularity of revealed data} & Implicit (Context only) $\leftrightarrow$ Explicit (Raw Signals) & Transparency $\leftrightarrow$ Anxiety/Nocebo     \\ \hline
2. Monitoring Temporality & \textbf{Timing of feedback}           & On-Demand $\leftrightarrow$ Continuous (Real-time)               & Intervention $\leftrightarrow$ Dependence  \\ \hline
3. Interpretation Framing & \textbf{Rhetorical mode of message}              & Reflective (Inquiry) $\leftrightarrow$ Suggestive (Directive)    & Autonomy $\leftrightarrow$ Cognitive Load  \\ \hline
4. AI Stance              & \textbf{Displayed confidence and role}         & Humble (Uncertain) $\leftrightarrow$ Authoritative (Objective)   & Safety $\leftrightarrow$ Overtrust        \\ \hline
5. Contestability         & \textbf{User's ability to correct}    & Fixed (Fact) $\leftrightarrow$ Correctable (Negotiable)          & Helplessness $\leftrightarrow$ Empowerment \\ \bottomrule
\end{tabular}
\end{table*}

\subsection{Data disclosure: what and how much to show}
Data disclosure determines the granularity of sensor-based information presented to users. While providing raw numbers and graphs increase transparency, it also heightens risks of over-interpretation and anxiety. In Push contexts where numbers are immediately interpreted as danger, disclosure may trigger nocebo-like reactions rather than increase trust~\cite{kerasidou2020artificial}. Therefore, the goal of disclosure should not be maximisation but contextual appropriateness. Depending on the user’s question (why, how) and current state (anxious, high arousal), and attributes of the condition (chronic or acute) the system should adjust combinations of implicit expressions (e.g., breathing prompts, rest suggestions), abstract indicators (recovery, load), and explicit metrics (heart rate, HRV)~\cite{amershi2019guidelines, singhal2023large}.

\subsection{Monitoring temporality: when to intervene}
Temporality is a lever between on-demand interaction (upon user request), episodic feedback (daily or weekly summaries), and continuous intervention (real-time alerts). Continuous monitoring can support early intervention, but it can also strengthen surveillance stress and dependence. If Push alerts occur for minor variations, notification fatigue and anxiety can accumulate, and users may experience the system as a monitor or supervisor. A safer default may therefore be on-demand or episodic designs, enabling continuous intervention only in higher-risk contexts and coupling it with safety mechanisms such as uncertainty marking, escalation, and contestability~\cite{singhal2023large, world2024ethics}.

\subsection{Interpretive framing: supporting reflection versus issuing directives}
Framing can make the same information lead to entirely different emotional and behavioural responses. Ambiguous sensor signals (e.g., elevated heart rate) are difficult to attribute to a single cause. Directive framing using controlling language (e.g., “You are stressed, so stop and rest now”) can increase anxiety or strengthen erroneous self-diagnosis. Reflective framing (e.g., “Was there exercise, caffeine, or a stressful meeting?”) helps users integrate context and co-construct meaning. Directive guidance is not always wrong for behaviour change, but given risks of overconfidence and hallucination in medical contexts~\cite{haug2023artificial, singhal2023large, kerasidou2020artificial}, directives should be conditional and verification-based, ideally accompanied by alternative hypotheses and user choice~\cite{amershi2019guidelines, lee2023benefits}.

\subsection{AI stance: how confidently, and with what role, the system speaks}
AI stance is not merely tone; it shapes whether users perceive the system as an authority resembling a medical professional. From clinical perspectives, LLMs can state incorrect information with confidence~\cite{lee2023benefits, haug2023artificial, singhal2023large, kerasidou2020artificial}, and healthcare discussions emphasise oversight needs~\cite{mesko2023imperative}. Excessive empathy can also lead to the empathy trap, strengthening overtrust and emotional dependence~\cite{mesko2023imperative, kerasidou2020artificial}. Therefore, sensor-fused health CAs should default to a humble and collaborative stance, reveal uncertainty, clarify role boundaries, and speak in ways that support user decision-making rather than replace it. In particular, controlling language such as “the sensor says” can intensify the illusion of objectivity, so systems should present evidence together with limitations~\cite{world2024ethics, singhal2023large, amershi2019guidelines}.

\subsection{Contestability}
From Correction to escalation, Sensor-fused inference is context-dependent, and users are often the first to detect mismatches. If a system continuously reifies incorrect labels (e.g., “stress”), users may feel their lived experience is denied. Contestability is a fundamental safety mechanism, not merely a UI option, enabling users to reflect, revise, or reject interpretations~\cite{amershi2019guidelines}. This dimension must extend to human-in-the-loop safeguards~\cite{singhal2023large}, where users can correct false positives or escalate high-risk alerts to professionals. Establishing these correction pathways prevents “learned helplessness” and clarifies role boundaries, ensuring the agent remains a support tool not an unquestionable authority.


\section{Risk Mitigation: Breaking the Biofeedback Loop}
Because sensor-fused health CAs are feedback dynmics/loops, the core of risk mitigation is designing interface guardrails that interrupt harmful biofeedback loops. Especially when Push notifications arrive during periods of health anxiety or high arousal, transparency can itself become harmful. What is needed is not “always show more,” but contextually appropriate disclosure and safe interaction transitions~\cite{kerasidou2020artificial, world2024ethics}. However, achieving this safety with static UIs is difficult because user readiness and vulnerability differ vastly between Pull and Push contexts. Therefore, adaptive interfaces that shift spectrum positions according to context become a necessary condition for safety~\cite{amershi2019guidelines}.

\subsection{Adaptive guardrails: automatic adjustment of expression intensity based on user state}
We propose guardrails that automatically shift selected dimensions of the design space toward the safer side when the system detects user states consistent with vulnerability (e.g., possible high arousal or anxiety). The key is a simultaneous switch: lowering disclosure to abstract forms, shifting AI stance toward humility, and moving framing toward reflection—replacing “you are stressed” with context-checking prompts (e.g., “heightened arousal detected; any recent exercise?”). This is not cosmetic UX refinement; it is safety design aimed at reducing overconfidence risks in healthcare LLMs~\cite{lee2023benefits, haug2023artificial, singhal2023large, kerasidou2020artificial} and the empathy trap~\cite{mesko2023imperative}.

\subsection{Verification-in-the-loop: uncertainty marking and short verification routines}
When LLMs provide medically sensitive advice, users must be able to receive it in a form that supports verification. Clinical discussions note that LLM risks often begin with a tone of certainty~\cite{haug2023artificial, singhal2023large, kerasidou2020artificial}. Therefore, the front end should implement short verification routines that combine uncertainty marking (probability, confidence, evidence), minimal context checking, and alternative hypotheses. For example, rather than fixing “stress” immediately, the system can first translate the signal more neutrally as “heightened arousal detected,” then ask the user to choose among “exercise, meeting, excitement, anxiety, other.” This allows low-cost self-report to be integrated before stabilising interpretation, preserving autonomy while handling contextual uncertainty.

\section{Conclusion}
For sensor-fused health CAs, front-end ethics is inseparable from care ethics. Design must move beyond model accuracy to focus on the users experience of biometrics translation. Although operationalizing principles remains challenging,
aligning interpretative transparency with global safety standards is essential. 
We hope this design space helps practitioners implement safeguards that protect autonomy, ensuring agents support rather than destabilize health.

\bibliographystyle{ACM-Reference-Format}
\bibliography{sample-base}

\end{document}